\documentclass[aps,prd,twocolumn,twoside,superscriptaddress,floatfix, nofootinbib]{revtex4-1}
\pdfoutput=1
\usepackage{graphicx}
\usepackage{color}
\usepackage{hyperref}
\usepackage{float}
\usepackage{aas_macros}
\usepackage{ifthen,amsmath,amssymb, bm}
\usepackage{multirow}
\usepackage{comment}

\graphicspath{{Figures/}}

\newcommand{\beq} {\begin{equation}}
\newcommand{\eeq} {\end{equation}}
\newcommand{\bal} {\begin{aligned}}
\newcommand{\eal} {\end{aligned}}

\newcommand{\am}[1]{{\textcolor{blue}{[AM: #1]}}}
\newcommand{\es}[1]{{\textcolor{green}{[ES: #1]}}}

\newcommand{\bk}{\boldsymbol{k}}
\newcommand{\bq}{\boldsymbol{q}}
\newcommand{\bv}{\boldsymbol{v}}
\newcommand{\bb}{\boldsymbol{\beta}}
\newcommand{\bx}{\boldsymbol{x}}

\newcommand{\barr}{\begin{eqnarray}}
\newcommand{\earr}{\end{eqnarray}}

\newcommand{\vsp}{\vphantom{\Big[}\\}

\begin{document}

\title{Doppler boosted dust emission and CIB-galaxy cross-correlations: a new probe of cosmology and astrophysics}
\author{Abhishek S. Maniyar}
\affiliation{Center for Cosmology and Particle Physics, Department of Physics, New York University, New York, NY 10003, USA}
\author{Simone Ferraro}
\affiliation{Lawrence Berkeley National Laboratory, One Cyclotron Road, Berkeley, CA 94720, USA}
\affiliation{Berkeley Center for Cosmological Physics, Department of Physics, University of California, Berkeley, CA
94720, USA}
\author{Emmanuel Schaan}
\affiliation{Lawrence Berkeley National Laboratory, One Cyclotron Road, Berkeley, CA 94720, USA}
\affiliation{Berkeley Center for Cosmological Physics, Department of Physics, University of California, Berkeley, CA
94720, USA}

\date{\today}

\begin{abstract}

We identify a new cosmological signal, the Doppler-boosted Cosmic Infrared Background (DB-CIB), arising from the peculiar motion of the galaxies whose thermal dust emission source the cosmic infrared background (CIB).
This new observable is an independent probe of the cosmic velocity field, highly analogous to the well-known kinematic Sunyaev-Zel'dovich (kSZ) effect.
Interestingly, DB-CIB does not suffer from the `kSZ optical depth degeneracy', making it immune from the complex astrophysics of galaxy formation.
We forecast that the DB-CIB effect is detectable in the cross-correlation of CCAT-Prime and DESI-like experiments.
We show that it also acts as a new CMB foreground which can bias future kSZ cross-correlations, if not properly accounted for.

\end{abstract}

\maketitle

\section{Introduction} \label{sec:intro}
The kinematic Sunyaev-Zel'dovich (kSZ) effect is the shift in the energy of the Cosmic Microwave Background (CMB) photons when they undergo Thomson scattering off coherently moving electrons in the gas in galaxies, groups, and clusters \cite{Sunyaev_1972, Sunyaev_1980}. The kSZ signal is linear in gas density and independent of the temperature of the gas. This makes it a crucial unbiased probe of these electrons on the outskirts of halos and clusters out to high redshift which are otherwise hard to detect. 

Using different techniques, the kSZ signal has been successfully measured through a combination of the CMB and galaxy survey data \citep[e.g.][]{Schaan_2016, Schaan_2021, Kusiak_2021}.
Thus, the kSZ effect is now a well-established tool to localize the ``missing baryons'' which reside outside the virial radius of the galaxies in an ionized, diffuse, and cold gas known as the warm-hot intergalactic medium \cite{Schaan_2021}. Apart from being a tracer of this gas, the kSZ signal is also a powerful probe of the radial velocities on large scales \citep[e.g.][]{Zhang_2010, Deutsch_2018, Smith_2018}. This makes the kSZ effect a probe of dark energy \cite{DeDeo_2005}, modified gravity \cite{Mueller_2015}, cosmic growth rate of structure \cite{Alonso_2016}, primordial non-Gaussianity of local type ($f_{\rm NL}$) \cite{Munchmeyer_2019} when used in combination with other matter tracers like galaxies. These techniques, however, suffer from the well-known problem of `kSZ--optical depth degeneracy' where the overall normalization of the electron profile in a halo is not known very well. So although we can measure the shape of the velocity power spectrum well with a combination of kSZ and galaxies, this degeneracy leads to an unknown overall normalization of the measured velocity field. 

In this paper, we present a new observable which is very analogous to the kSZ effect but does not suffer from the `optical depth degeneracy' which kSZ suffers from. This observable is the Doppler boosted emission of the cosmic infrared background (CIB), which we will call DB-CIB from here onward. The CIB the is cumulative infrared emission from all the dusty star forming galaxies throughout the Universe \cite{Planck_cib_14}. It is an excellent probe of the cosmic star formation and the large scale structure of the Universe \cite{Maniyar_2018, Maniyar_2019}. 

If a galaxy contributing to the CIB has a non-zero line-of-sight peculiar velocity, its emission is Doppler boosted.
The large-scale cosmic velocity field results in galaxy bulk motions, which in turn source the DB-CIB signal of interest in this paper.

Unlike kSZ, the DB-CIB does not originate from scattering CMB photons:
the Doppler boosting is imprinted on the galaxy's thermal dust emission.
However, the DB-CIB signal is deeply analogous to the kSZ: it measures the product of velocities with the mean infrared luminosity.
Crucially, this mean infrared luminosity can be measured independently, unlike the kSZ optical depth.
It is thus `calibratable' and can be removed, providing unbiased estimates of the velocity field. 
This is precisely the reason we do not have an analogous optical depth degeneracy here.
In this paper, we compute for the first time the expected signal-to-noise (SNR) for the detection of this signal for a Planck-like \cite{Planck_specs_2014} and Fred Young submillimeter telescope (CCAT-Prime) -like \cite{CCAT_2020} experiment. This effect also acts as a contaminant to the kSZ measurements from the CMB power spectrum and from the cross-correlation of CMB with galaxies. We will quantify this contamination in this work. 

The remainder of this paper is organized as follows. In the following Sec.~\ref{sec:doppler}, we derive the formalism to quantify this DB-CIB emission. Then in Sec.~\ref{sec:cross}, we present the formalism to detect this effect through cross-correlation of the CIB with velocity-weighted density field. We then present the expected SNR of this signal for the Planck and CCAT-Prime experiments in combination with the CMASS \cite{SDSS_2013} catalog from the
Baryon Oscillation Spectroscopic Survey (BOSS) and the Dark Energy Spectroscopic Instrument (DESI) galaxy survey. In Sec.~\ref{sec:contamination} and \ref{sec:contamination_auto}, we calculate the expected contamination from this signal to kSZ measurements, while we present potential applications of this signal in Sec.~\ref{sec:discuss}.

\section{Doppler-boosted CIB emission} \label{sec:doppler}
For any specific intensity $I(\nu)$ at frequency $\nu$, the quantity $I(\nu)/\nu^3$ is a conserved quantity under Lorentz transformations, including boosts.
Neglecting the cosmological expansion for now, 
the frequency of a source moving towards us gets boosted such that $\nu_{\rm 0} = D_+ \nu_{\rm em}$ where $\nu_{\rm 0}$ and $\nu_{\rm em}$ are observed and emitted frequencies respectively, and $D_+ = \gamma(1 + \beta \cos{(\theta)}) \approx 1 + \beta$ for radial motion and $\beta \ll 1$ ($\beta = v/c$ with $v$ being the source velocity with respect to us and $c$ is the speed of light), and $\gamma = 1/\sqrt{1-\beta^2}$ is the usual Lorentz factor. We neglect any transverse component of the Doppler effect because it is second order in $\beta$. 
Taking cosmological expansion into account, $\nu_{\rm 0} = D_+ \nu_{\rm em}/(1+z) \approx (1+\beta) \nu_{\rm em}/(1+z)$ where $z$ is the source redshift. Using the above-mentioned Lorentz invariance, we obtain
\beq
\label{eq:lorentz}
\frac{I^{\rm obs}(\nu_0)}{\nu_0^3} = \frac{I^{\rm RF}(\frac{\nu_0(1+z)}{(1+\beta)})}{\left(\frac{\nu_0(1+z)}{(1+\beta)}\right)^3} \approx (1+3\beta) \frac{I^{\rm RF}(\frac{\nu_0(1+z)}{(1+\beta)})}{(\nu_0(1+z))^3} \, ,
\eeq
where $I^{\rm obs}(\nu)$ and $I^{\rm RF}(\nu)$ are the observed and rest frame specific intensities respectively. Denoting $\nu'_{\rm em} = \nu_0(1+z)$, we can Taylor expand $I^{\rm RF}(\frac{\nu_0(1+z)}{(1+\beta)})$ 
\beq
\label{eq:irf_taylor}
I^{\rm RF}\Big(\frac{\nu'_{\rm em}}{1+\beta}\Big) \approx I^{\rm RF}(\nu'_{\rm em}) - \nu'_{\rm em} \beta \frac{dI^{\rm RF}(\nu'_{\rm em})}{d\nu'_{\rm em}} \, .
\eeq
Thus, the fractional change in the specific intensity due to Doppler boosting is
\barr
\frac{\Delta I(\nu_0)}{I(\nu_0)} 
&\equiv& \frac{I^{\rm obs}(\nu_0) - I^{\rm obs}(\nu_0)|_{\beta=0}}{I^{\rm obs}(\nu_0)|_{\beta=0}} \nonumber \\
&=& \frac{(1+3\beta) \left(I^{\rm RF}(\nu'_{\rm em}) - \nu'_{\rm em} \beta \frac{dI^{\rm RF}(\nu'_{\rm em})}{d\nu'_{\rm em}} \right) - I^{\rm RF}(\nu'_{\rm em})}{I^{\rm RF}(\nu'_{\rm em})} \nonumber\\
&=&
\beta
\left(
3 - \underbrace{\frac{d\ln I^{\rm RF}(\nu'_{\rm em})}{d\ln \nu'_{\rm em}}}_\alpha
\right)
+\mathcal{O}\left( \beta^2 \right) \nonumber\\
&=&
\beta 
\left(
3 - \alpha
\right)
+\mathcal{O}\left( \beta^2 \right).
\label{eq:doppler_single_galaxy_frequency}
\earr
Here, we introduced the the logarithmic slope of the observed intensity with respect to the observed frequency $\alpha$ (see Eq.~\ref{eq:alpha_rest_observed}).
The equation above gives us the relative size of the DB-CIB and the usual CIB effects.
The DB-CIB is reduced by a factor the typical line-of-sight velocity $\beta = v/c \sim 10^{-3}$.
The $3-\alpha$ factor indicates that the DB-CIB effect is absent if the specific intensity $I^{\rm RF}$ scales as the cube of the frequency, since such a specific intensity is invariant under Doppler boosts.

Using the fact that $I(\nu)/\nu^3$ is Lorentz invariant, we get from Eq.~\eqref{eq:lorentz} for $\beta=0$:
\beq \label{eq:alpharestobs}
I^{\rm RF}(\nu'_{\rm em}) = (1+z)^3 I^{\rm obs}(\nu_0 = \nu'_{\rm em}/(1+z)) \, ,
\eeq
%
%
such that we can simply replace:
\beq
\frac{d \ln I^{\rm RF}(\nu'_{\rm em})}
{d\ln \nu'_{\rm em}}
=
\frac{d\ln I^{\rm obs}}{d\ln \nu_0} (\nu_0 = \nu_\text{em}'/(1+z)).
\label{eq:alpha_rest_observed}
\eeq

Since the dust emission from galaxies is typically observed within a frequency bandpass $W(\nu_0)$, we need to integrate the above equation over our given bandpass.
Hence the observed signal:
\beq
\bal
\label{eq:dInu_Inu_band}
\frac{\Delta I_W}{I_W} 
&=\frac{\int \left[ I^{\rm obs}(\nu_0)  - I^{\rm obs}(\nu_0)|_{\beta=0}\right] W(\nu_0) d\nu_0}{\int I^{\rm obs}(\nu_0)|_{\beta=0} W(\nu_0) d\nu_0}\\
&=
\beta
\left( 3 -
\underbrace{\frac{\int 
\frac{d \ln I^{\rm obs}}{d\ln \nu_0} I^{\rm obs}(\nu_0)|_{\beta=0}
W(\nu_0) d\nu_0}
{\int I^{\rm obs}(\nu_0)|_{\beta=0} W(\nu_0) d\nu_0}}
_{\alpha_{\rm filt}}
\right)
+\mathcal{O}\left( \beta^2 \right)\\
&=
\beta(3-\alpha_{\rm filt}) + \mathcal{O}\left( \beta^2 \right)
\eal
\eeq
In particular, for a narrow tophat bandpass, this indeed simplifies to Eq.~\eqref{eq:doppler_single_galaxy_frequency}. 
Importantly, all the equations so far apply to \textit{any} emission process, including infrared emission giving rise to the CIB, but also synchrotron emission and any other radiative process across the whole electromagnetic spectrum.

\begin{figure}[h]
\centering
\includegraphics[width=\columnwidth]{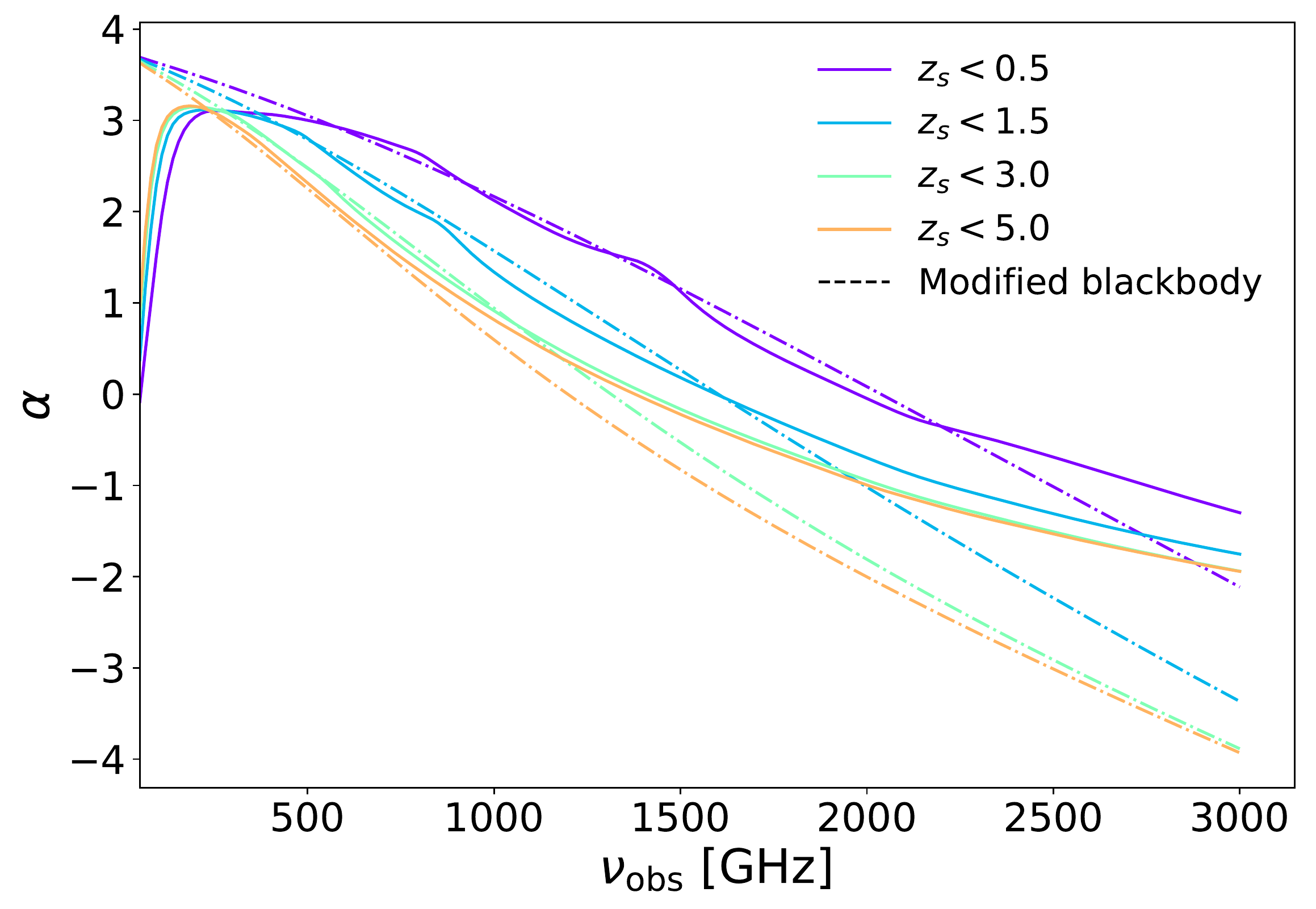}
\centering 
\caption{$\alpha$ as a function of the observed frequency and redshift. Different colors correspond to the source galaxies  contributing to the CIB coming from redshifts between 0 and $z_s$ i.e. ($0 < z < z_s$). The dashed lines correspond to the modified blackbody choice of the effective SEDs where $\beta_d=1.75$ and $T_d = T_0 (1+z)^\gamma$ with $T_0 = 24.4$ K and $\gamma=0.36$ from \cite{Planck_cib_14}.
}
\label{fig:alpha_general}
\end{figure}

In what follows, we shall study the case of the CIB in detail, since the CIB dominates the extragalactic emission at millimiter and sub-millimeter frequencies. For the CIB, $I^{\rm obs}(\nu_0)$ can be calculated using Eq.~\eqref{eq:inu}, details of which are provided in Appendix \ref{app:halo}. This requires a prior knowledge of the effective spectral energy distribution (SED) $S^{\rm eff}_{\nu_0}(z)$ of the Infrared (IR) galaxies at a given redshift and frequency. We use the $S^{\rm eff}_{\nu_0}(z)$ templates from a stacking analysis presented in \cite{Bethermin_2015}. 
An alternative approach in \cite{Planck_cib_14} fits for $S^{\rm eff}_{\nu_0}(z)$ with a modified blackbody parameterization such that $S^{\rm eff}_{\nu_0}(z) \propto \nu_0^{\beta_d} B_{\nu_0}(T_d(z))$ where $B_{\nu_0}$ denotes the Planck function, $T_d$ denotes the dust temperature as a function of redshift, and $\beta_d$ is the emissivity index encoding information about the physical nature of the dust. In Fig.~\ref{fig:alpha_general}, we show $\alpha$ as a function of the observed frequency and redshift for these two choices of SEDs.   
Looking at Eqs. \eqref{eq:doppler_single_galaxy_frequency} or \eqref{eq:dInu_Inu_band}, we see that the DB-CIB emission is proportional to a factor of $(3 - \alpha)$. We find that $\alpha \approx 3$ for frequencies between $\sim$ 100-500 GHz for different redshifts when we use SEDs from \cite{Bethermin_2015}. Thus, the Doppler-boosted signal might be reduced for these choices of frequencies. 
Interestingly, the spectral index $\alpha$ becomes negative at high frequencies. 
This indicates a drop-off of intensity with respect to the observed frequency, when observing above 1.5-2.5 THz depending on the model and the redshift of the source. Since the factor of $(3-\alpha)$ is increased for negative $\alpha$, the Doppler boosting of the CIB more prominent at higher frequencies.

At very low frequencies ($\nu < 70$ GHz) where we expect the CIB intensity to drop-off, template SEDs from \cite{Bethermin_2015} instead flatten out, leading to $\alpha \approx 0$ in this case. At such low frequencies, synchrotron radiation coming from extragalactic sources compensates for the drop in the infrared emission making the final intensity almost constant with frequency which results in $\alpha \approx 0$. While this effect is included in the template SEDs from \cite{Bethermin_2015}, it is not included in the modified blackbody template shown in the dashed curves and therefore the value of $\alpha$ differs between the two SEDs at these low frequencies. The synchrotron radiation itself is also Doppler boosted, allowing us to treat it with the same formalism.




\section{Cross-correlation with galaxies} \label{sec:cross}
Our approach for detection of the DB-CIB signal follows the kSZ detections by \cite{Schaan_2016, Schaan_2021}, who stacked the ACT CMB maps, appropriately weighted by an external tracer of peculiar velocity, at the positions of the BOSS galaxies.

Here we suggest a similar procedure to detect the DB-CIB emission signal through cross-correlation of the observed CIB field with a density-weighted velocity field (momentum) $\bq(\bx)$ from the galaxy positions. We work in Fourier space for convenience, and we assume that an external template of the velocity $\bb$ at the galaxy positions is available, for example from using the continuity equation (as in baryonic acoustic oscillation (BAO) reconstruction), as explained in Section III.B of \cite{Schaan_2021}.

To predict the signal, we shall adopt the formalism of Ref.~\cite{Ma_2002}, who provided fully nonlinear expressions for the temperature fluctuations from the kSZ effect. Since we expect the DB-CIB to matter only on small scales ($\ell \gg 100$), we shall use their high-$\ell$ approximation throughout. 

Let us denote the galaxy momentum field $\bq(\bx) \equiv (1+\delta_g)\bb$ where $\delta_g$ is the galaxy overdensity and thus
\beq
\tilde{\bq}(\bk) = \tilde{\bb}(\bk) + \int \frac{d^3k'}{(2\pi)^3} \tilde{\bb}(\bk')\tilde{\delta}_g(\bk - \bk') \, ,
\label{eq:q(k)}
\eeq
where $\tilde{\bb}$ and $\tilde{\delta}_g$ are Fourier transforms of $\bb$ and $\delta_g$ respectively. We will rely on the cross-correlation of the CIB fluctuations $\Delta I(\nu_0)$ with the $\bq(\bx)$ field to detect the Doppler-boosted signal, in analogy with previous kSZ work. Specifically, we will be cross-correlating $\Delta I(\nu_0)$ with $\bq_\gamma$ which is the line of sight component of $\bq(\bx)$ i.e. $\hat{\gamma} \cdot \bq$. From Eqs.~\eqref{eq:doppler_single_galaxy_frequency} and \eqref{eq:alpha_rest_observed}, we can see that $\Delta I(\nu_0) = (3-\alpha) I(\nu_0) \beta$.
In the Limber and flat sky approximation, the angular cross-power spectrum of $\Delta I(\nu_0)$ and $\bq_\gamma(\bx)$ is given as
\beq
C^{\Delta I_{\nu_0} q_\gamma}_\ell = \int \frac{dz}{\chi^2} \frac{d\chi}{dz} \: W^{\Delta I_{\nu_0}}(\nu_0, z) \: W^{q}(z) P^{\Delta I_{\nu_0} q_\gamma}\left(\frac{\ell}{\chi}, z\right) \, ,
\label{eq:cell_cibdoppler}
\eeq
where $\chi$ is the comoving distance to redshift $z$, $W^{\Delta I_{\nu_0}}(\nu_0, z)$ and $W^{q}(z)$ are the window functions corresponding to the CIB fluctuations and the galaxy survey, respectively. $P^{\Delta I_{\nu_0} q_\gamma}(\frac{\ell}{\chi}, z)$ is the cross-power spectrum of the CIB fluctuations and the line-of-sight component $\hat{\gamma} \cdot \tilde{\bq}$, which we compute next.

Now, $\tilde{\bq}(\bk)$ can be divided into a longitudinal mode $\tilde{\bq}_\parallel(\bk) = \left( \tilde{\bq}\cdot\hat{\bk} \right)\hat{\bk}$ and a transverse mode $\tilde{\bq}_\perp(\bk) = \tilde{\bq} - \left( \tilde{\bq}\cdot\hat{\bk} \right) \hat{\bk}$ on the plane parallel and perpendicular to the mode vector $\bk$ -- not the line of sight -- such that $\tilde{\bq} = \tilde{\bq}_\parallel + \tilde{\bq}_\perp$. 
We  focus on small angular scales where the flat sky and Limber approximations are valid.
In this regime, the Limber line-of-sight integral selects only the Fourier modes with 
$\bk\cdot\hat{\gamma} = 0$.
For these Fourier modes, the line-of-sight momentum which sources the DB-CIB thus simplifies to
$\tilde{\bq}(\bk)\cdot\hat{\gamma} 
=
\left([
\tilde{\bq}_\parallel(\bk)
+
\tilde{\bq}_\perp(\bk)
\right]
\cdot \hat{\gamma}
=
\tilde{\bq}_\perp(\bk)
\cdot \hat{\gamma}.
$
As a result, $P^{q_\gamma}(k, z) = \frac{1}{2} P^{q_\perp}(k, z)$ for these Fourier modes $\bk$ perpendicular to the line of sight.
Following the same reasoning, we also obtain:

\beq
P^{\Delta I_{\nu_0} q_\gamma}(k, z) = \frac{1}{2} P^{\Delta I_{\nu_0} q_\perp}(k, z) \, .
\label{eq:approx_perp}
\eeq
Following the calculations in \cite{Ma_2002} (Eqs.~5, 6, and 8), in the high-$k$ limit, we get:
\beq
P^{\Delta I_{\nu_0} q_\perp}(k, z) \approx \frac{2 \left( 3-\alpha \right)}{3} P^{\Delta I_{\nu_0} \delta_g}(k, z) \int \frac{d^3 k'}{(2\pi)^3}P^{\beta \beta}(k', z)  \, ,
\label{eq:power_highk}
\eeq
where $P^{\Delta I_{\nu_0} \delta_g}(k, z)$ is the cross-power spectrum of the CIB fluctuations and the galaxy overdensity field. 

We recognize the integral in the equation above as being related to the variance of the line of sight velocity:
\beq
\bal
\langle \beta_\text{LOS}^2 \rangle(z)
&\equiv
\frac{1}{3}\int \frac{d^3k'}{(2\pi)^3} P^{\beta \beta}(k')\\
&=
\frac{1}{3} \left(\frac{a H(z) f }{c}\right)^2
\int \frac{d^3k'}{(2\pi)^3} \frac{1}{k'^2}P^{\delta \delta}(k', z)
\label{eq:beta_rms}
\eal
\eeq
Here $H$ is the Hubble parameter, $f$ is the linear growth factor, which is well approximated by $f \approx \Omega_m(z)^\gamma$ where $\Omega_m(z)$ is the matter density at redshift $z$ and $\gamma \approx 0.55$ in General Relativity. Therefore, Eq.~\eqref{eq:power_highk} becomes
\beq
P^{\Delta I_{\nu_0} q_\gamma}(k, z) = \left( 3-\alpha \right) P^{\Delta I_{\nu_0} \delta_g}(k, z) \langle \beta_\text{LOS}^2 \rangle(z) \, .
\label{eq:kSZ_fin}
\eeq

Eq. \ref{eq:kSZ_fin}, together with Eq. \ref{eq:cell_cibdoppler} and the approximation in Eq. \ref{eq:approx_perp} represent the main result of this paper. The last ingredient needed to evaluate the expected signal is the cross-correlation between CIB fluctuations and galaxies, $P^{\Delta I_{\nu_0} \delta_g}(k, z)$. We calculate this cross power spectrum following the CIB halo model from \cite{Maniyar_21} and the details are provided in Appendix \ref{app:halo}.

\section{Forecasts} \label{subs:snr}

Here we present the signal-to-noise (SNR) ratio for the $C^{\Delta I_{\nu_0} q_\gamma}_\ell$ cross-correlation, which is the signal we are after. The SNR is calculated as
\beq
\left( \frac{S}{N} \right)^2 = f_{\rm sky} \sum_{{\ell_b}_{\rm min}}^{{\ell_b}_{\rm max}}  \frac{(2{\ell_b}+1)  \Delta \ell \left(C_{\ell_b}^{\Delta I_{\nu_0} q_\gamma}\right)^2} {\left(C_{\ell_b}^{\Delta I_{\nu_0} q_\gamma}\right)^2 + C_{\ell_b}^{\Delta I_{\nu_0} \Delta I_{\nu_0}} \times C_{\ell_b}^{q_\gamma q_\gamma}} \, ,
\label{eq:snr}
\eeq
where 
\beq
C_{\ell_b} = \frac{1}{\Delta \ell} \sum_{\ell \in [\ell_1, \ell_2]} C_\ell \, ,
\eeq
and $\Delta \ell$ is the bin width. 

$C_{\ell_b}^{\Delta I_{\nu_0} \Delta I_{\nu_0}}$ is the total binned CIB auto power spectrum at frequency $\nu_0$ i.e. it is the sum of the one-halo, two-halo and the shot-noise power spectra, $C_{\ell_b}^{\Delta I_{\nu_0}\Delta I_{\nu_0}} = C_{\ell_b, 1h}^{\Delta I_{\nu_0}\Delta I_{\nu_0}} + C_{\ell_b, 2h}^{\Delta I_{\nu_0}\Delta I_{\nu_0}} + C_{\ell_b, {\rm shot}}^{\Delta I_{\nu_0} \Delta I_{\nu_0}}$. To evaluate this term, we also add a detector white-noise term $N_\ell^\mathrm{det}$ for various experiments described below. 

$C_{\ell_b}^{q_\gamma q_\gamma}$ is the galaxy radial velocity field power spectrum.
It is obtained from 
$P^{q_\gamma q_\gamma}(k, z)=P^{\delta_g \delta_g}(k, z)\langle \beta_\text{LOS}^2 \rangle(z)$, following Eq.~\eqref{eq:kSZ_fin}. 

As previously mentioned, we use a halo model approach to calculate all the auto- and cross-power spectra, with full details in App.~\ref{app:halo}. 
Similar to the case of the CIB, for the galaxy auto- and CIB $\times$ galaxy cross-power spectra, we sum up the 1-halo, 2-halo, and shot noise power spectrum contributions. For the CIB $\times$ galaxy power spectra, we estimate the cross-shot noise term for a given frequency as
\begin{equation} \label{eq:crossshot}
    C_{\ell_b, {\rm shot}}^{\Delta I_{\nu_0}\delta_g } = \sqrt{C_{\ell_b, {\rm shot}}^{\Delta I_{\nu_0} \Delta I_{\nu_0}} \times C_{\ell_b, {\rm shot}}^{\delta_g \delta_g}} \, .
\end{equation}
In practice, this is an upper limit to the cross-shot noise term, as it assumes that the shot noise of CIB and galaxies are perfectly correlated.
Since the actual level of cross-shot noise is uncertain, we only include it when forecasting the CIB SNR, not the DB-CIB SNR. 
Perhaps counterintuively, this choice is actually conservative, and can only lead to underestimating the DB-CIB SNR.
Indeed, the cross-shot noise is both part of our signal and noise (via its cosmic variance), but the noise contribution is negligible, since we are far from cosmic variance limited.
Formally, this can be seen from Eq.~\eqref{eq:snr}, where the cross-shot noise term appears both in the numerator and the denominator.
However, in the denominator, the cross-power spectrum is small compared to the product of the auto spectra, in our noise dominated regime.
As a result, including the cross-shot noise would have no effect on the noise, but would enhance the signal. 
This enhanced signal will mostly be seen on very small scales ($\ell \gtrsim 3000$).

For the CIB part, we assume two different setups which correspond to Planck-like and CCAT-Prime-like experiments. For the galaxy survey, we assume four different galaxy samples corresponding to the CMASS-like ($0.44 < z < 0.70$), DESI-ELG-like ($0.0 < z < 2.0$), DESI-LRG-like ($0.0 < z < 1.4$), and extended DESI-ELG-like ($0.0 < z < 4.0$ and denoted as Ext.~DESI-ELG) galaxy samples. 
Ext.~DESI-ELG is assumed to be a hypothetical galaxy survey which detects the same number of galaxies 
as DESI-ELG survey, but extended over twice the redshift range. 
To calculate the galaxy and CIB $\times$ galaxy power spectra within a halo model framework, a halo occupation distribution (HOD) is required. Here we use the HOD corresponding to the CMASS survey developed by \cite{More_2015}. We use the same HOD parametrization for the DESI-ELG, DESI-LRG, and Ext.~DESI-ELG samples as well with a minor tweak: we adjust the minimum galaxy mass detectable for different samples such that the total numbers of galaxies detected by these surveys match the expected numbers from these surveys. While not exact, this should be a reasonable approximation for our purposes. The sky fraction is assumed to be $f_{\rm sky} = 0.4$.  

Our assumed experimental setups which correspond to the Planck-like and CCAT-Prime-like experiments are given in Tab.~\ref{tab:exp}. The Gaussian random noise of the detector is calculated as 
\begin{equation}
    N_\ell^\mathrm{det} = (\Delta_{T})^2 e^{\ell(\ell + 1) \sigma^2 /8 \ln{2}}
\end{equation}
where $\Delta_{T}$ denotes the white noise of the detector in $\mu$K-arcmin or Jy/sr, and $\sigma$ is the full width at half maximum (FWHM) of the beam in radians. 
As shown in \cite{Maniyar_2019, Lenz_2019}, galactic dust dominates over the CIB power spectra below $\ell \sim 100$, and therefore we choose $\ell_{\rm min} = 100$ in Tab.~\ref{tab:exp}.

\begin{table}[h]
\begin{tabular}{|c|c|c|c|c|c}
\hline
Experiment & $\ell_\mathrm{min}$ & $\ell_\mathrm{max}$ & $\Delta_T$ & $\sigma$ \vsp
 & & & $\mu\mathrm{K}$-arcmin & arcmin \vsp
\hline
Planck (545 GHz) & 100 & 5000 & 1137.0 & 4.7  \vsp
Planck (857 GHz) & 100 & 5000 & 29075.0 & 4.3  \vsp
CCAT-prime (410 GHz) & 100 & 50000 & 372.0 & 0.5  \vsp
CCAT-prime (850 GHz) & 100 & 50000 & 5.7$\times 10^5$ & 0.2  \vsp
\hline
\end{tabular}
\centering \caption{Experimental specifications used in this work. Planck and CCAT-Prime specifications are taken from \cite{Planck_specs_2014} and \cite{CCAT_2020}. The detector noise is quoted in thermodynamic differential CMB temperature units.}
\label{tab:exp}
\end{table}

Our predictions for the expected SNR on the $C_{\ell}^{\Delta I_{\nu_0} \delta_g}$ and $C_{\ell}^{\Delta I_{\nu_0} q_\gamma}$ for the experimental specifications considered here are given in Tab.~\ref{tab:SNR_cibxg} and \ref{tab:SNR_cibxv} respectively. As an in-depth study of the flux-cut limits for CCAT-Prime is beyond the scope of this paper, we consider two limiting cases: (i) shot noise for CCAT-Prime is equal to the shot-noise for Planck and (ii) CCAT-Prime has 10 times lower shot-noise than Planck. While the 3rd column in Tab.~\ref{tab:SNR_cibxg} and \ref{tab:SNR_cibxv} corresponds to case (i), the 4th column corresponds to case (ii) for CCAT-Prime experiment.

\begin{table}[h]
\begin{tabular}{|c|c|c|c|}
\hline
CIB exp & Galaxy exp & \multicolumn{2}{|c|}{SNR} \vsp
        &            &   High shot & Low shot \\
\hline
{Planck} & CMASS & 1065 (1430) & \\ 
\cline{2-3}
545 (857) & DESI-ELG & 1370 (1868) & \\
GHz & DESI-LRG & 1216 (1666) & \\
\cline{2-3}
 & Ext.~DESI-ELG & 1483 (1871) & \\
\hline
{CCAT-Prime} & CMASS & 8907 (6453) & 4118 (2397) \\
\cline{2-4}
410 (850) & DESI-ELG & 10043 (8186) & 6073 (4357) \\ 
GHz & DESI-LRG & 9377 (7191) & 4912 (3220) \\ 
\cline{2-4}
 & Ext.~DESI-ELG & 10935 (8646) & 7506 (4848) \\
\hline
\end{tabular}
\centering \caption{SNR for CIB $\times$ galaxy power spectra for different configurations considered here. For CCAT-Prime-like experiment considered here, in the 3rd column we assume shot noise to be equal to shot noise from Planck for corresponding frequency, while in the 4th column shot noise for CCAT-Prime is 10 times smaller than Planck. Unbracketed and bracketed numbers show SNR at 545 and 857 GHz respectively for Planck, and at 410 and 850 GHz respectively for CCAT-Prime.}
\label{tab:SNR_cibxg}
\end{table}

As can be seen from these tables, $C_{\ell}^{\Delta I_{\nu_0} \delta_g}$ can be detected to very high SNR. On the other hand, the DB-CIB signal $C_{\ell}^{\Delta I_{\nu_0} q_\gamma}$ detection will be challenging with a Planck-like experiment considered here. A combination of CCAT-prime and DESI surveys should be able to detect $C_{\ell}^{\Delta I_{\nu_0} q_\gamma}$ with a high ($> 5$) SNR for 850 GHz channel. 

\begin{table}[h]
\begin{tabular}{|c|c|c|c|c|}
\hline
CIB exp & Galaxy exp & \multicolumn{2}{|c|}{SNR} \vsp
        &            &   High shot & Low shot \\
\hline
{Planck} & CMASS & 0.05 (0.37) & \\ 
\cline{2-3}
545 (857)  & DESI-ELG & 0.98 (5.03) & \\
GHz & DESI-LRG & 0.35 (3.66) & \\
\cline{2-3}
 & Ext.~DESI-ELG & 1.75 (5.67) & \\
\hline
{CCAT-Prime } & CMASS & 0.01 (2.30) & 0.01 (2.35) \\
\cline{2-4}
410 (850) & DESI-ELG & 3.71 (51.77) & 4.36 (52.82) \\ 
GHz & DESI-LRG & 1.96 (31.27) & 2.32 (31.93) \\ 
\cline{2-4}
 & Ext.~DESI-ELG & 15.21 (68.42) & 18.00 (69.85) \\
\hline
\end{tabular}
\centering \caption{SNR for $C^{\Delta I_{\nu_0} q_\gamma}_\ell$ different different configurations considered here. For CCAT-Prime-like experiment considered here, in the 3rd column we assume shot noise to be equal to shot noise from Planck for corresponding frequency, while in the 4th column shot noise for CCAT-Prime is 10 times smaller than Planck. Unbracketed and bracketed numbers show SNR at 545 and 857 GHz respectively for Planck, and at 410 and 850 GHz respectively for CCAT-Prime.}
\label{tab:SNR_cibxv}
\end{table}
The CMASS, DESI, and Ext.~DESI surveys considered here trace galaxies around redshifts $\sim 0.5$, $\sim 1.0$, and $\sim 2.0$ respectively. Therefore, SNR for $C_{\ell}^{\Delta I_{\nu_0} \delta_g}$ for Planck is higher with 857 GHz channel than 545 GHz channel as for the CIB higher frequencies trace relatively lower redshifts and vice-versa \citep[e.g.][]{Maniyar_2018}. However, this is not the case for CCAT-Prime experiment where SNR is lower for 850 GHz than 410 GHz. This is mainly due to the significant higher instrumental noise at 850 GHz than 410 GHz. The logic applied here has to be slightly modified while looking at Tab.~\ref{tab:SNR_cibxv} for SNR on $C_{\ell}^{\Delta I_{\nu_0} q_\gamma}$. As we can see from Eq.~\eqref{eq:kSZ_fin}, calculation of $C_{\ell}^{\Delta I_{\nu_0} q_\gamma}$ from $C_{\ell}^{\Delta I_{\nu_0} \delta_g}$ involves extra factors of $(3 - \alpha)$ and $\langle \beta_\text{LOS}^2 \rangle(z)$ which depend on frequency and redshift respectively. The factor of $(3 - \alpha)$ is smaller at 545 (410) GHz than at 857 (850) GHz (Fig.~\ref{fig:alpha_general}). Also, for the redshifts considered here, $\langle \beta_\text{LOS}^2 \rangle(z)$ decreases with increasing redshifts. Combining these two things again with the fact that CIB at higher frequencies traces galaxies at lower redshifts, we can see that SNR for $C_{\ell}^{\Delta I_{\nu_0} \delta_g}$ and $C_{\ell}^{\Delta I_{\nu_0} q_\gamma}$ is higher at 857 (or 850) GHz than at 545 (or 410) GHz for Planck (or CCAT-Prime) experiment considered here. We note that in this calculation we use the actual redshift range corresponding to our galaxy samples to calculate $\alpha$ unlike what we show in  Fig.~\ref{fig:alpha_general}, where $\alpha$ is calculated after integrating the CIB emission between $0 < z < z_s$ for different source redshifts $z_s$.

For the extended DESI-ELG like survey considered here, we see that both $C_{\ell}^{\Delta I_{\nu_0} \delta_g}$ and $C_{\ell}^{\Delta I_{\nu_0} q_\gamma}$ are detected at higher SNR than other surveys. 
In the case of $C_{\ell}^{\Delta I_{\nu_0} \delta_g}$ this is solely due to obtaining the signal over a larger range of redshift (thus larger overlap with CIB redshifts \cite{Maniyar_2018}) compared to other surveys. As can be seen from Fig.~\ref{fig:alpha_general}, the value of $\alpha$ is lower when galaxies over a broad redshift range (e.g.~Ext.~DESI-ELG: $0 < z < 4$) are considered compared to a narrower range (e.g. DESI-ELG: $0 < z < 2$). This results in a higher value of the $(3 - \alpha)$ factor which enters in the calculation of $C_{\ell}^{\Delta I_{\nu_0} q_\gamma}$ for surveys of broader redshift range. Combined with the larger redshift overlap with the CIB, this effect adds to have higher SNR for $C_{\ell}^{\Delta I_{\nu_0} q_\gamma}$ detection with Ext.~DESI-ELG survey compared to other surveys.

The SNR in case (ii) for CCAT-Prime experiment in Tab.~\ref{tab:SNR_cibxg} is smaller than in case (i) which has a higher shot noise compared to former. This is because the cross-shot noise term given in Eq.~\ref{eq:crossshot}, adds to the signal for CIB $\times$ galaxy cross-correlation. This is not the case for $C_{\ell}^{\Delta I_{\nu_0} q_\gamma}$ where there is no cross-shot noise term in Eq.~\ref{eq:snr} and only the auto-shot power spectrum for the CIB and galaxy survey appear in the denominator acting as noise decreasing the SNR. 
Therefore, unlike for CIB $\times$ galaxy, the SNR slightly increases for case (ii) compared to case (i). 
In other words, unlike for the case of CMB observations where decreasing foreground levels by masking sources is beneficial, in our case the Doppler-boosted emission from the sources \textit{is} our signal, and therefore aggressive masking is not guaranteed to lead to higher SNR. In fact, more aggressive masking will reduce the noise (by reducing shot noise), but will also reduce the signal. A full study of the optimal flux cuts that maximize the SNR is beyond the scope of this paper.

\section{Contamination to kSZ measurements from Doppler-boosted CIB} \label{sec:contamination}
As pointed out in the introduction, the kSZ effect has been measured in several analyses to date. Most estimators\footnote{With the notable exception of the ``projected fields'' estimators \cite{Ferraro:2016ymw, Hill:2016dta, Kusiak_2021}.} rely on the fact that the imprint of kSZ on the CMB maps is correlated with the galaxies' peculiar velocities and can be extracted by cross-correlating a template for the peculiar velocities with the CMB maps themselves. This has an additional advantage: any foreground contamination to the small-scale CMB which is uncorrelated with the peculiar velocity vanishes on average. This is true for the bulk of the CIB emission or the thermal SZ signal: while larger than kSZ in amplitude, they cancel in the kSZ estimator, allowing for kSZ extraction from single-frequency maps.
In this paper, we have pointed out that due to Doppler boosting, part of the CIB emission \textit{is} correlated with the peculiar velocity, and hence will not cancel in the kSZ estimator and will act as a bias to kSZ measurements.

In App.~\ref{appsubs:doppler_single}, we calculate the bias at CMB frequencies from galaxies of a given mass and redshift. For example, for BOSS CMASS galaxies with mean halo mass $M_h \approx 10^{13} M_\odot$ (\cite{Schaan_2021}) at $z \sim 0.5$, we find that the DB-CIB contamination in the CMB maps is $\approx 3\times10^{-4}, 7\times10^{-4}, 2\times10^{-3} \: \mu {\rm K}$ at 100, 143, and 217 GHz respectively, when averaged over a disk with radius $= 0.5$ arcmin.
The kSZ effect on the same aperture is of the order of $0.01 - 0.1 \mu {\rm K}$ \cite{Schaan_2016, Schaan_2021}, and hence the Doppler-boosted bias is one to two orders of magnitude lower than the signal. For the sensitivity levels of the current CMB maps, this effect can thus be neglected. However, experiments like Simons Observatory and CMB-S4 will measure the kSZ signal with SNR larger than 100 \cite{Battaglia:2017neq, SimonsObservatory:2018koc, S4_19}. For such more sensitive experiments, this contamination will have to be considered to get unbiased estimates of the kSZ signal.

For a given CIB SED model, the frequency dependence of the DB-CIB is approximately known, such that it can be extracted or nulled via internal linear combinations (ILC) of multi-frequency maps. However, this process suffers from the same uncertainties in the CIB SED modeling as when extracting CIB maps from multi-frequency data. In techniques where stacking of the CMB data at the locations of galaxies is used to detect the kSZ \citep[e.g.][]{Schaan_2021}, 
the CIB emission profile is found to be pointlike (i.e. unresolved), and therefore much more concentrated than the kSZ signal. 
Indeed, the CIB emission originates from the dust inside the galaxy ($\sim 10$~kpc), whereas kSZ is sourced by the gas profile, much more extended ($\sim 1$~Mpc).
(Fig.~11 of \cite{Schaan_2021}). 
Therefore, in such analyses, we can exclude the data in this narrow region around the center of the stacked image to avoid contamination from the DB-CIB emission. While this will reduce the dominant part of the 1-halo term, it will not remove the contamination from other galaxies correlated with the sample in question (whether in the same halo or not). 

\section{The Doppler-boosted CIB as an additional small-scale CMB anisotropy}
\label{sec:contamination_auto}
In addition to the bias to kSZ discussed in the previous Section, we note that this signal is also an extra small-scale CMB anisotropy with its own specific power spectrum and frequency dependence. 

This extra anisotropy may contaminate searches for the kSZ power spectrum, which in turn can be used to study the epoch of reionization.
CMB data between $\sim 100-200$ GHz is generally used for such purposes. At these frequencies, contamination from the CIB is usually an issue and \cite{Maniyar_21} and has to be removed in order to obtain unbiased measurements of the kSZ power spectrum. The DB-CIB emission presented here will also act as an extra source of foreground emission for the kSZ. 

In App.~\ref{appsubs:doppler}, we present a formalism to calculate the power spectrum of the DB-CIB emission. Using, Eq.~\eqref{eq:doppler_auto}, we predict the value of the power spectrum $\frac{\ell (\ell+1)}{2\pi} C_{\ell, D}^{\Delta I_{\nu_0} \Delta I_{\nu_0}}$ where `D' in the subscript denotes that power spectrum is for the DB-CIB component. We find that at $\ell = 3000$, the amplitude is $\sim 10^{-9}, 10^{-8}, 10^{-7} \: \mu {\rm K}^2$ at 100, 143, and 217 GHz respectively. This should be compared to the (frequency-independent) value for kSZ at $\ell = 3000$ of $\frac{\ell (\ell+1)}{2\pi} C_{\ell}^{\rm kSZ} \sim 3 \ \mu {\rm K}^2$ \cite{Shaw:2011sy}, a value that is rather uncertain due to several astrophysical unknowns needed to predict it. 

Overall, this shows that the DB-CIB power spectrum is predicted to be negligible compared to the kSZ power spectrum around $\ell = 3000$. Thus it appears that the DB-CIB signal will not be a major contaminant to the detection of the kSZ power spectrum.

As mentioned in the Sec.~\ref{subs:snr}, as we go higher in frequency, CIB traces galaxies at lower redshifts to some extent where value of the peculiar velocity $\beta(z)$ is higher than at high redshifts. Also, the CIB power spectrum increases with frequency. These two effects result in the value of the power spectrum of the DB-CIB emission to increase with frequency. 


\section{Discussion and Conclusions} \label{sec:discuss}
Emission coming from the CIB galaxies gets boosted by the Doppler effect as a result of their motion in the large scale cosmological velocity field. 
In this paper, we present a formalism to calculate this effect
and quantify the detectability of the
cross-correlation of the CIB with a velocity weighted galaxy density field. 
We show that although this effect would be hard to detect through a cross-correlation of Planck and CMASS/DESI galaxies, a combination of the CCAT-Prime and DESI survey can potentially detect this signal. 

We also show this effect constitutes a new source of foreground while measuring the kSZ power spectrum, and a bias to stacking-based kSZ estimators. For upcoming CMB experiments like SO and CMB-S4 which plan to detect the kSZ at a very high significance, this foreground contamination will have to be considered and removed. We point out in Sec.~\ref{sec:contamination}, this can be done using the distinct frequency dependence, as well as the different angular profile of this effect.

As mentioned in Sec.~\ref{sec:intro}, the radial velocity field is an excellent cosmological probe. It has been shown that the kSZ tomography technique can be successfully used to measure the radial velocity field with the upcoming CMB surveys \cite{Smith_2018, Munchmeyer_2019}. Due to the `kSZ optical depth degeneracy', the overall normalization of the measured velocity is not known a priory and must be marginalized over. 
This is not an issue for measurements of $f_{\rm NL}$ due to the scale dependence of the signal, but it poses a significant challenge for measurements that require knowledge of the normalization, such as  growth of structure which depend on the amplitude of the velocity power spectrum. 

From Eq.~\eqref{eq:doppler_single_galaxy_frequency}, we can see that the DB-CIB emission can act a new observable to reconstruct the velocity field $\beta$, free from this degeneracy. Thus, we can construct an estimator for $\beta$ using a combination of a CIB and galaxy survey or solely using the CIB. This estimator has an advantage over the kSZ tomography technique as it does not suffer from the `optical depth degeneracy' as the intensity of the CIB emission at a give frequency $I_\nu$ is \textit{calibratable} by direct measurement of the cross-correlation $C^{\Delta I_{\nu_0} \delta_g}_\ell$ (or by stacking). As can be seen from Table \ref{tab:SNR_cibxg}, the SNR on $C^{\Delta I_{\nu_0} \delta_g}_\ell$ is always a lot greater the SNR of the DB-CIB signal, so that the uncertainty on the calibration is always subdominant and should not limit the inference of the velocity field.

Thus, the velocities detected through such a technique will be a useful cosmological probe. In fact, it has to be noted that such an effect of Doppler boosting is not limited to the CIB emitting galaxies and is generalizable to any galaxy population. Therefore, such a formulation can be used with the galaxies detected through the powerful upcoming surveys like DESI, Euclid and Roman Space Telescope. In an upcoming paper, we will present such an estimator of velocity and its predictions for cosmological constraints.

\acknowledgments

We thank Jacques Delabrouille, Colin Hill, Anthony Pullen and David Spergel for useful conversations. SF is supported by the Physics Division of Lawrence Berkeley National Laboratory.
ES is supported by the Chamberlain fellowship at Lawrence Berkeley National Laboratory.

\bibliographystyle{prsty.bst}
\bibliography{refs}

\onecolumngrid
\appendix

\section{CIB, galaxy, CIB$\times$galaxy halo power spectrum} \label{app:halo}
\subsection{CIB Power spectrum}
The angular power spectrum of the CIB anisotropies is defined as
\begin{equation} \label{eq:clgen}
\Big \langle \delta I_{\ell m}^{\nu} \delta I_{\ell'm'}^{\nu'} \Big \rangle = C_\ell^{\Delta I_\nu \Delta I_{\nu' }} \times \delta _{\ell \ell'}\delta_{mm'} \, ,
\end{equation}
where $\nu$ is the frequency of the observation and $I_\nu$ is the specific intensity of the CIB measured at that frequency. 

The specific intensity is a function of the comoving emissivity $j$ through
\begin{equation} \label{eq:inu}
\begin{split}
I_\nu & =  \int \dfrac{d\chi}{dz} a j(\nu, z) dz \\
 & =  \int \dfrac{d\chi}{dz} a \bar{j}(\nu, z) \Big( 1 + \frac{\delta j(\nu, z)}{\bar{j}(\nu, z)} \Big) dz \, ,
\end {split}
\end{equation}
where $\chi(z)$ is the comoving distance to redshift $z$, and $a = 1/(1+z)$ is the scale factor of the Universe. Combining Eqs. \ref{eq:clgen} and \ref{eq:inu}, and using the Limber's approximation (Limber et al. 1954), we get
\begin{equation} \label{eq:cl_lin2h}
C_\ell^{\Delta I_\nu \Delta I_{\nu' }} = \int \frac{dz}{\chi^2} \frac{d\chi}{dz}a^2 \bar{j}(\nu,z) \bar{j}(\nu',z)P^{\Delta I_\nu \Delta I_{\nu' }}(k = \ell/\chi, z) \, ,
\end{equation}
where $P^{\Delta I_\nu \Delta I_{\nu' }}$ is the 3D power spectrum of the emissivity and is defined as
\begin{equation}
\Big \langle \delta j(k,\nu) \delta j(k',\nu') \Big \rangle = (2\pi)^3 \bar{j}(\nu)\bar{j}(\nu')P^{\Delta I_\nu \Delta I_{\nu' }}(k)\delta^3(k-k')\, ,
\end{equation}
where $\delta j$ are the emissivity fluctuations of the CIB. Under the assumption that the CIB is sourced by the galaxies, the emissivity power spectrum can be equated with the galaxy power spectrum. \\
We proceed through integrating over the specific emissivities of the halos at a given frequency rather than using a luminosity to halo mass parameterization as is done normally. In the halo model formalism we assume that there is a central galaxy at the center of the halo and the satellite galaxies are occupying the satellite halos which follow the NFW distribution. The 1-halo term is given as
\begin{align}
\label{eq:1halo}
P_{1h}^{\Delta I_\nu \Delta I_{\nu' }}(k, z) &= \frac{1}{\overline{j}_\nu\overline{j}_{\nu'}} \int  \Bigg[ \frac{dj_{\nu,c}}{d\log M_h} \frac{dj_{\nu',sub}}{d\log M_h}u(k,M_h,z) + \frac{dj_{\nu',c}}{d\log M_h} \frac{dj_{\nu,sub}}{d\log M_h}u(k,M_h,z) \notag \\
&+ \frac{dj_{\nu,sub}}{d\log M_h} \frac{dj_{\nu',sub}}{d\log M_h}u^2(k,M_h,z)\Bigg]\: {\left( \frac{d^2N}{d\log M_hdV} \right)} \, d\log M_h
\end{align}
where $\frac{d^2N}{d\log M_hdV}$ is the halo-mass function \cite{Tinker_2008}, $u(k,M_h,z)$ is the Fourier transform of the NFW profile describing the density distribution inside the halo and $\frac{dj}{d\log M_hdV}(\nu,z)$ is the differential emissivity of the central and satellite subhalos at a given frequency and redshift for a given halo mass. \\
Specific emissivity is given as
\beq
\frac{dj_{\nu,c}}{d\log M_h}(M_h,z) = \chi^2 (1+z) \times \frac{\mathrm{SFR}_c}{K} \times S^{\rm eff}_\nu(z) 
\eeq
where $S^{\rm eff}_\nu(z)$ is the effective SED of the IR galaxies at a given redshift for a given frequency. We use the $S^{\rm eff}_\nu(z)$ templates computed using a stacking analysis presented in \cite{Bethermin_2015}. $\mathrm{SFR}_c$ is the star formation rate for the central galaxies with a given halo mass. $K$ is the Kennicutt constant $K = \mathrm{SFR}/L_{IR} = 1 \times 10^{-10} M_\odot\mathrm{yr}^{-1}{\mathrm{L}_\odot}^{-1}$ for a Chabrier IMF with $L_{IR}$ being the IR luminosity. \\
For the satellite galaxies
\beq
\frac{dj_{\nu,sub}}{d\log M_h}(M_h,z) = \chi^2 (1+z) \times S^{\rm eff}_\nu(z) \int \frac{dN}{d\log m_\mathrm{sub}}(m_\mathrm{sub}|M_h) \frac{\mathrm{SFR}_\mathrm{sub}}{K}\: d\log m_\mathrm{sub}
\eeq
where $\frac{dN}{d\log m_\mathrm{sub}}$ is the sub-halo mass function for the satellite galaxies with a sub-halo mass $m_\mathrm{sub}$ \cite{Tinker_2010}. The effective SEDs $S^{\rm eff}_\nu(z)$ for the satellite galaxies are assumed to be the same as the central galaxies. $\mathrm{SFR}_\mathrm{sub}$ is the star formation rate for the satellite galaxies with a given subhalo mass. 
Total emissivity is given as:
\beq
\label{eq:jnu}
j_\nu = \int \Bigg( \frac{dj_{\nu, c}}{d\log M_h} + \frac{dj_{\nu, sub}}{d\log M_h}\Bigg) \times \frac{d^2N}{d\log M_hdV} \, d\log M_h
\eeq
The 2-halo term is given as
\barr
\label{eq:2halo}
P_{2h}^{\Delta I_\nu \Delta I_{\nu' }}(k, z) &=& \frac{P_\mathrm{lin}(k, z)}{\overline{j}_\nu\overline{j}_\nu'} \times \int \Bigg[ \frac{dj_{\nu, c}}{d\log M_h} + \frac{dj_{\nu, sub}}{d\log M_h}u(k,M_h,z)\Bigg] \, b(M_h, z) \frac{d^2N}{d\log M_hdV} \, d\log M_h \nonumber \\
&& \times \int  \Bigg[ \frac{dj_{\nu',c}}{d\log M'_h} + \frac{dj_{\nu', sub}}{d\log M'_h}u(k,M_h,z)\Bigg] \, b(M'_h, z) \frac{d^2N}{d\log M'_hdV} \, d\log M'_h
\earr
where $b(M_h, z)$ is the halo bias \cite{Tinker_2010_b} and $P_\mathrm{lin}(k, z)$ is the linear dark matter power spectrum.

We compute the CIB power spectrum using the halo model presented in \cite{Maniyar_21}. Details of the calculations of the CIB power spectrum are presented in Appendix \ref{app:halo}. Using Eq.~\ref{eq:1halo} and \ref{eq:2halo} in Eq.~\ref{eq:cl_lin2h}, we can get $C^{\Delta I_\nu \Delta I_\nu'}_{\ell, 1h}$ and $C^{\Delta I_\nu \Delta I_\nu'}_{\ell, 2h}$ with $k = \ell/\chi$. 

\subsection{Galaxy power spectrum}
Following \cite{Cooray_02}, the 1-halo and 2-halo terms of the galaxy power spectrum are
\begin{equation}
\label{eq:1halogal}
P^{\delta_g}_{1h}(k, z) =  \int \frac{d^2N}{d\log M_hdV} \frac{2N_c \, N_s u(k, M_h, z) + N_s^2 \, u^2(k, M_h, z)}{{\overline{n}}^2_\mathrm{gal}} \, d\log M_h \, ,
\end{equation}
and
\begin{equation}
\label{eq:2halogal}
P^{\delta_g}_{2h}(k, z) =  P_\mathrm{lin}(k, z) \Bigg\{ \int \frac{d^2N}{d\log M_hdV} \frac{\bigg(N_c + N_s \, u(k, M_h, z)\bigg) \times b(M_h, z)}{{\overline{n}}_\mathrm{gal}} \, d\log M_h \Bigg\}^2 \, ,
\end{equation}
where 
\begin{equation}
\overline{n}_\mathrm{gal}(z) = \int \frac{d^2N}{d\log M_hdV} (N_c + N_s) \, d\log M_h
\end{equation}
where $N_c(M_h, z)$ and $N_s(M_h, z)$ are the number of central and satellite galaxies inside a given halo respectively which are given by the HOD model. 

If we assume that all the galaxies equally contribute to the CIB i.e. there is no dependence of the luminosity or the emissivity of a halo on the halo mass, it can be checked that Eq.~\ref{eq:1halo} and \ref{eq:2halo} reduce to Eq. \ref{eq:1halogal} and \ref{eq:2halogal}. The corresponding power spectra in the harmonic space $C_\ell$ can be obtained using Eq.~\ref{eq:cl_lin2h} by substituting $\overline{j}(\nu)$ with $\overline{n}_\mathrm{gal}$.

\subsection{CIB $\times$ galaxy power spectrum}
Finally using Eq.~\ref{eq:1halo}, \ref{eq:2halo}, \ref{eq:1halogal}, and \ref{eq:2halogal} the CIB$\times$galaxy cross-correlation is given as

\barr
\label{eq:1halocross}
P^{\Delta I_\nu \delta_g}_{1h}(k, z) = \frac{1}{\overline{n}_\mathrm{gal} \overline{j}(\nu)} \int \frac{d^2N}{d\log M_hdV} \Bigg(N_c + N_s \, u(k, M_h, z)\Bigg) \Bigg( \frac{dj_{\nu, c}}{d\log M_h} + \frac{dj_{\nu, sub}}{d\log M_h}u(k,M_h,z)\Bigg) \, d\log M_h \, ,
\earr
and
\barr
\label{eq:2halocross}
P^{\Delta I_\nu \delta_g}_{2h}(k, z) = && \frac{P_\mathrm{lin}(k, z)}{\overline{n}_\mathrm{gal} \overline{j}(\nu)} \int \frac{d^2N}{d\log M_hdV} b(M_h, z) \bigg(N_c + N_s \, u(k, M_h, z)\bigg) \, d\log M_h \nonumber \\
&& \times \int \frac{d^2N}{d\log M'_hdV} b(M'_h, z) \bigg( \frac{dj_{\nu,c}}{d\log M'_h} + \frac{dj_{\nu, sub}}{d\log M'_h}u(k,M'_h,z)\bigg) \, d\log M'_h  \, .
\earr

Following Eqs.~\ref{eq:1halocross} and \ref{eq:2halocross}, we get
\barr
\label{eq:halocross_cl}
C^{\Delta I_\nu \delta_g}_{\ell, 1h(2h)} &=& \int \frac{dz}{\chi^2} \frac{d\chi}{dz} \: W^{\Delta I_\nu}(\nu, z) \: W^{\delta_g}(z) P^{\Delta I_\nu \delta_g}_{1h(2h)} (k=\ell/\chi, z) \, ,
\earr
where $W^{\Delta I_\nu}(\nu, z)$ and $W^{\delta_g}(z)$ are the window functions for the CIB and galaxy survey we are considering respectively, and $P^{\Delta I_\nu \delta_g}_{1h(2h)}(k=\ell/\chi, z)$ is the CIB$\times$galaxy 1 and 2-halo cross power spectrum given by Eqs.~\ref{eq:1halocross} and \ref{eq:2halocross}.  
The window functions are given as
\beq
W^{\Delta I_\nu}(\nu, z) = a \times j(\nu, z)
\eeq
\beq
W^{\delta_g}(z) = \frac{dz}{d\chi} \times \frac{dp}{dz} = \frac{dz}{d\chi} \times \frac{dN/dz}{\int dz \frac{dN}{dz}}
\eeq
where $j$ is given using Eq.~\ref{eq:jnu} and $\frac{dN}{dz}$ is the number of galaxies within a given redshift interval for the survey we are considering. 

\subsection{Doppler boosted power spectrum} \label{appsubs:doppler}
Here we present the formulation to calculate the power spectrum of only the Doppler boosted component of the CIB emission $C_{\ell, D}^{\Delta I_{\nu} \Delta I_{\nu'}}$. In Sec.~\ref{sec:cross}, we follow the derivation of the nonlinear kSZ power spectrum $C^{q_\gamma q_\gamma}_\ell$ from \cite{Ma_2002} to calculate $C^{\Delta I_{\nu} q_\gamma}_\ell$. We do this by cross-correlating the line of sight velocity weighted density field $q_\gamma$ with the DB-CIB emission $\Delta I_{\nu}$. As a result we obtain Eq.~\eqref{eq:kSZ_fin}. We can follow the similar procedure from \cite{Ma_2002} but for the auto-power spectrum of the line of sight velocity weighted CIB field. Similar to Eq.~\eqref{eq:kSZ_fin}, in the high-k nonlinear regime, we get 
\beq
P_{D}^{\Delta I_{\nu} \Delta I_{\nu'}}(k, z) = \left( 3-\alpha \right) \left( 3-\alpha' \right) P^{\Delta I_{\nu} \Delta I_{\nu'}}(k, z) \langle \beta_\text{LOS}^2 \rangle(z) \, ,
\label{eq:doppler_fin}
\eeq
where D denotes the DB-CIB power spectrum and $\alpha$ and $\alpha'$ are calculated at frequencies $\nu$ and $\nu'$ respectively. Substituting this in Eq.~\eqref{eq:cl_lin2h}, we get
\barr \label{eq:doppler_auto}
C_{\ell, D}^{\Delta I_{\nu} \Delta I_{\nu'}} &=& \int \frac{dz}{\chi^2} \frac{d\chi}{dz}a^2 \bar{j}(\nu,z) \bar{j}(\nu',z)P_{D}^{\Delta I_{\nu} \Delta I_{\nu'}}(k = \ell/\chi, z) \nonumber \\
&=& \int \frac{dz}{\chi^2} \frac{d\chi}{dz}a^2 \left( 3-\alpha \right) \left( 3-\alpha' \right) \bar{j}(\nu, z) \bar{j}(\nu', z) \langle \beta_\text{LOS}^2 \rangle(z) P^{\Delta I_{\nu} \Delta I_{\nu'}}(k = \ell/\chi, z) \, .
\earr

\subsection{Doppler boosted emission from a single halo} \label{appsubs:doppler_single}
Here we present the formalism to calculate the DB-CIB emission ($\Delta I(\nu)$) coming from a single halo with a mass $M$ at redshift $z$. The CIB formalism presented here deals with a population of CIB halos at a given redshift. This makes the calculation for a single halo a bit tricky. Thus, for this estimate, we will rely on the luminosity-halo mass relation $L-M$ used in the other CIB models \citep[e.g.][]{Planck_cib_14, Fiona_2021}. This relation is given as 
\beq
L_{(1+z)\nu}(M, z) = L_0 \Phi(z)\Sigma(M)\Theta\left[(1+z)\nu \right] \, ,
\eeq
where
$L_0$ is an overall normalisation factor; $\Sigma(M)$ gives the mass dependence; $\Phi(z)$ provides the redshift evolution of the $L-M$ relation; and $\Theta(\nu)$ is the SED of the galaxy considered. Overall parameterizations for these factors and the best-fit values of the corresponding parameters after fitting to the CIB power spectrum data are given in \cite{Fiona_2021} which we use here. 

\begin{figure}[h]
\centering
\includegraphics[width=0.8\columnwidth]{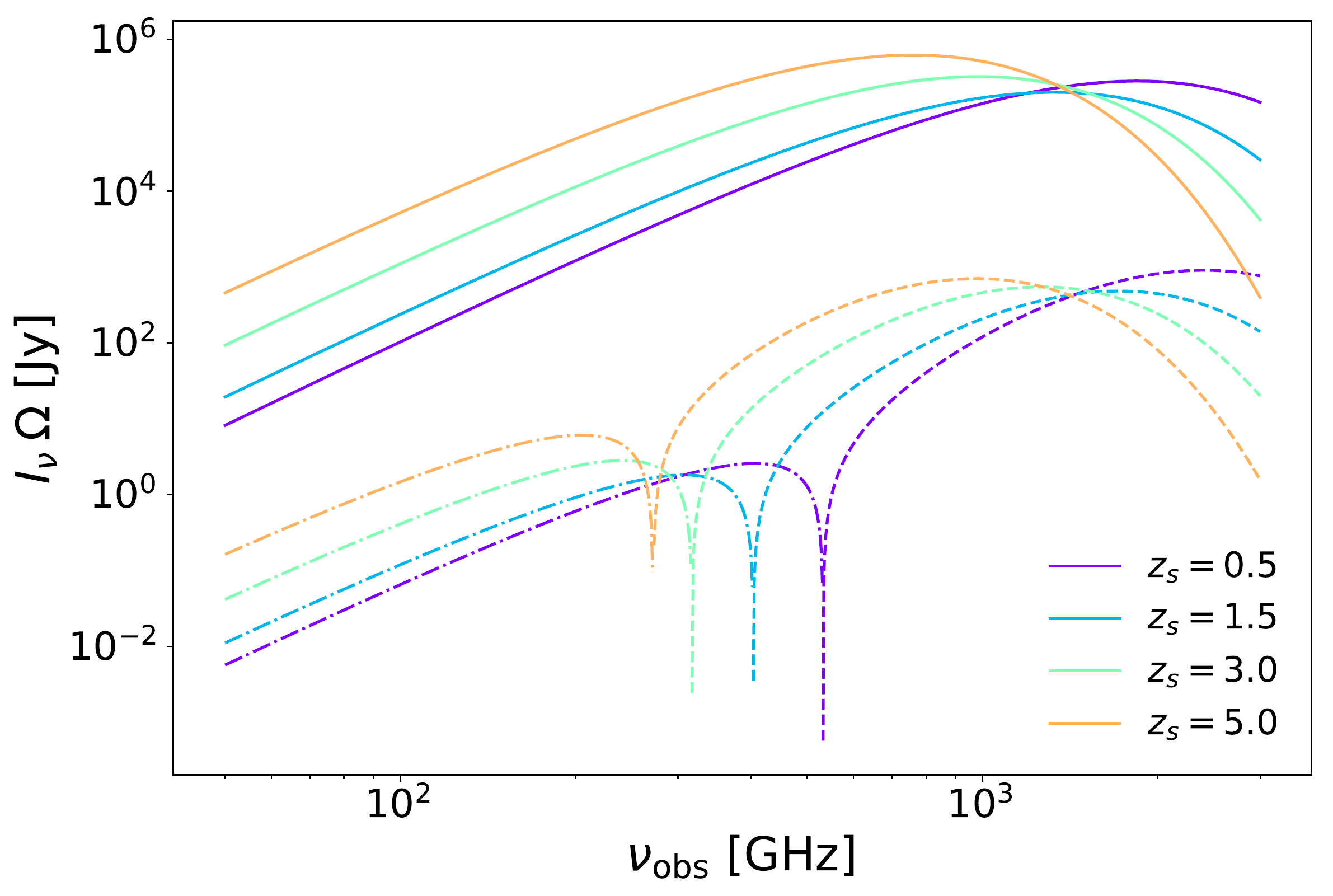}
\centering 
\caption{Solid lines show $I(\nu) \Omega$ calculated using Eq.~\eqref{eq:Inu_modblack} for a single halo of mass $10^{13} M_\odot$ at various redshifts. 
The solid lines show the usual, non-Doppler-boosted thermal dust emission from the halo.
Dashed and dot-dashed lines show $\Delta I(\nu) \Omega > 0$ and $\Delta I(\nu) \Omega < 0$ calculated using Eq.~\eqref{eq:deltaInumodblack} respectively. $\Delta I(\nu) \Omega$ becomes negative at low frequencies because for the modified blackbody SED for the galaxy which we use to create this plot, $\alpha > 3$ (and thus $(3-\alpha) < 0$) at low frequencies as shown in Fig.~\ref{fig:alpha_general}. The modified blackbody template for the SED of the galaxy does not include the synchroton emission which a galaxy might emit at low frequencies, and therefore the behavior of an individual galaxy might change based on its emission properties at these low frequencies. }
\label{fig:Inu_deltaInu_signelgal}
\end{figure}

Once we have the luminosity of the galaxy, we can calculate the specific intensity (averaged over a galaxy subtending a solid angle $\Omega$) as
\beq \label{eq:Inu_modblack}
I(\nu) = \frac{L_{(1+z)\nu}}{4\pi \chi^2(z) (1+z) \Omega} \, ,
\eeq
where $\chi$ is the comoving distance for the galaxy redshift $z$. If the galaxy subtends an angle of $\theta$ at the detector, the solid angle is $\Omega = A/\chi^2(z)$ where $A=\pi (\chi(z)\theta)^2$, and therefore $\Omega = \pi \theta^2$. Finally from Eq.~\eqref{eq:doppler_single_galaxy_frequency}, the amplitude of the Doppler effect is
\beq \label{eq:deltaInumodblack}
\Delta I(\nu) = \beta(z) \left( 3 - \alpha \right) I(\nu) \, .
\eeq 
It has to be noted that in reality both $L_{(1+z)\nu}$ and $I(\nu)$ are dependent on the density profile of the galaxies within a halo (normally assumed to be NFW). We neglect that effect here assuming that the profile is constant with radius of the halo. In Fig.~\ref{fig:Inu_deltaInu_signelgal}, we show expected specific intensity and the corresponding DB-CIB emission multiplied by the solid angle factor for a single galaxy of mass $10^{13} M_\odot$ at various redshifts.

\end{document}